\documentclass{article}
\date{}

\usepackage{booktabs} 
\usepackage{pdflscape}
\usepackage{afterpage}
\usepackage{capt-of}

\usepackage{xcolor}

\usepackage[utf8]{inputenc}
\usepackage{flushend}

\usepackage{hyperref} 

\usepackage[export]{adjustbox}
\usepackage{fancybox}

\usepackage{array}

\newcolumntype{R}[2]{%
    >{\adjustbox{angle=#1,lap=\width-(#2)}\bgroup}%
    l%
    <{\egroup}%
}

\usepackage{mathtools}
\DeclarePairedDelimiter{\ceil}{\lceil}{\rceil}

\usepackage{enumitem}
\newenvironment{pil}{
\begin{itemize}[leftmargin=*]
  \setlength{\itemsep}{1pt}
  \setlength{\parskip}{0pt}
  \setlength{\parsep}{0pt}
}{\end{itemize}}

\usepackage{listings}
\definecolor{DarkGreen}{rgb}{0,0.3,0}
\definecolor{DarkBlue}{rgb}{0,0,0.7}
\definecolor{grey}{rgb}{0.5,0.5,0.5}
\definecolor{redOrange}{rgb}{0.8,0.2,0}

\lstdefinelanguage{Spectra}[]{}{
  commentstyle=\color{DarkGreen}\itshape,
  keywordstyle=[1]\color{blue}\bfseries,
  keywordstyle=[2]\color{DarkGreen}\bfseries,
  keywordstyle=[3]\color{red}\bfseries,
  keywordstyle=[4]\color{purple}\bfseries,
  morekeywords=[1]{G,GF,next,H,ONCE,boolean,false,true,spec,alw,always,alwaysEventually,alwEv,trans,ini,initially},
  morekeywords=[2]{out,sys,gar},
  morekeywords=[3]{in,env,asm},
  morekeywords=[4]{counter,overflow,inc,reset,underflow,type,monitor,define,predicate},
  morecomment=[l]{//},
  morecomment=[s]{/**}{*/}
}

\newtheorem{definition}{Definition}

\newcommand{\true}{\text{\small{\op{true}}}}
\newcommand{\op}[1]{\texttt{\color{blue}\bfseries{#1}}}

\lstdefinelanguage{grammar}[]{}{
  frameshape={RYRY}{y}{y}{RYRY},
  commentstyle=\color{DarkGreen}\itshape,
  keywordstyle=[1]\color{redOrange}\bfseries,
  keywordstyle=[5]\color{purple}\bfseries,
  morekeywords=[1]{spec,name,specElem,assumption,guarantee,varDec,type,
    exp,unaryOp,binaryOp,primExp,enumType,intType,intBinaryOp,
    intUnaryOp,def,defRef,typeDef,typeRef,pattern,varDecl,patInst,monitor,
    monRef,pLTLUnaryOp,pLTLBinaryOp,predicate,typedParam,predInst,file,
    intRef,enumValRef,int,patVar,specWithImports,intVal,tempConstraint,
    stateInvConstraint},
  morekeywords=[5]{extends,replaces},
  morecomment=[l]{//},
  morecomment=[s]{/**}{*/},
  literate={<|}{{$\langle$}}1 {|>}{{$\rangle$}}1 {::=}{{$::=$}}1
}
\newcommand{\key}[1]{\textbf{\color{blue}{#1}}}
\newcommand{\synt}[1]{\textbf{\color{grey}{#1}}}
\newcommand{\prodRule}[1]{$\langle$\texttt{\color{redOrange}{\bfseries{#1}}}$\rangle$}

\begin{document}

\title{Spectra:\\A Specification Language for Reactive Systems}

\author{Shahar Maoz\\
		School of Computer Science, Tel Aviv University, Israel \and
		Jan Oliver Ringert \\
		Department of Informatics, University of Leicester, UK}

\maketitle

\begin{abstract}

Spectra is a new specification language for reactive systems, specifically
tailored for the context of reactive synthesis. The meaning of Spectra is
defined by a translation to a kernel language. Spectra comes with the Spectra
Tools, a set of analyses, including a synthesizer to obtain a
correct-by-construction implementation, several means for executing the
resulting controller, and additional analyses aimed at helping engineers write
higher-quality specifications.
We present the language and give an overview of the tool set.  

\end{abstract}

\section{Introduction}

Reactive synthesis is an automated procedure to obtain a correct-by-construction
reactive system from its temporal logic specification~\cite{PR89}. Rather than
manually constructing an implementation and using model checking to verify it
against a specification, synthesis offers an approach where a correct
implementation of the system is automatically obtained for a given
specification, if such an implementation exists.

As the correct-by-construction promise is attractive, much research effort has
been invested and much progress has been achieved over the last two decades on
reactive synthesis theory, algorithms, and tools. These include, e.g., the
identification of expressive fragments of temporal logics that have efficient,
symbolic synthesis solutions~\cite{BJP+12,KupfermanLVY11,KupfermanV10}, various
kinds of compositional synthesis~\cite{LustigV13}, controller synthesis for
probabilistic systems~\cite{DragerFKPU14,KwiatkowskaP13}, support for the
addition of quantitative criteria~\cite{BloemCHJ09}, bounded
synthesis~\cite{FinkbeinerS13}, and tools such as RATSY~\cite{BloemCGHKRSS10}
and Slugs~\cite{EhlersR16}, to list a few. Still, major challenges remain on the
way to bringing the correct-by-construction promise to software engineering
practice.

Some of these challenges are at the level of the specification language, i.e.,
the syntax, semantics, and expressiveness of reactive specifications.   First,
the language should have features that are specific to its use in the context of
reactive synthesis, such as the explicit distinction between system and
environment variables, guarantees, and assumptions.  Second, careful balance
should be found between the language's usability to engineers, in writing and
reading, on the one hand, and its formal expressive power on the other hand.
Finally, the language alone may not be enough to address all challenges. Rather,
a set of analyses and tools, beyond synthesis itself, specifically tailored for
the new language and its use in an end-to-end reactive synthesis environment, is
required, for example, in executing the synthesized controllers and in debugging
them.  In this paper we describe the results of our research efforts over the
last three years, to start developing such language and tool set.

Spectra is a new specification language for reactive systems, specifically
tailored for the context of reactive synthesis. The meaning of Spectra is
defined by a translation to a kernel language.
Beyond the kernel, it includes advanced language features like patterns,
monitors, bounded integers, and arithmetic operations.
Spectra comes with the Spectra Tools, a set of analyses, including a synthesizer
to obtain a correct-by-construction implementation, several means for executing
the resulting controller, and additional analyses aimed at helping engineers
write higher-quality specifications, including, e.g., dealing with unrealizable and
non-well-separated specifications.

The paper is structured as follows.  Sect.~\ref{sec:background} provides
necessary background on linear temporal logic and its GR(1) fragment, which are
required for the formal definition of Spectra semantics. Sect.~\ref{sec:example}
presents an example Spectra specification of an autonomous forklift robot (first
reported in~\cite{MaozR15synt}). This specification illustrates many language
features of Spectra and serves as a running example throughout the paper.
Sect.~\ref{sec:kernel} introduces a small language kernel of Spectra and defines
the semantics of a specification as a GR(1) synthesis problem.
Sect.~\ref{sec:langElements} covers the language elements Spectra provides and
defines their semantics via translations to the kernel. In Sect.~\ref{sec:tools}
we give an overview of Spectra Tools.

\section{Background}
\label{sec:background}

We provide background on linear temporal logic and synthesis, which is required
for the definition of Spectra syntax and semantic.  We continue with the
notation that we use to define the grammar of Spectra.

\subsection{Linear Temporal Logic and Synthesis}

We repeat some of the standard definitions of linear temporal logic
(LTL), e.g., as found in~\cite{BJP+12}, a modal temporal logic with
modalities referring to time. The syntax of LTL formulas is
typically defined over a set of atomic propositions $\mathit{AP}$ with
the future temporal operators \op{X} (next) and \op{U} (until).

\vspace{-.5em}
\begin{definition} The syntax of LTL formulas over $\mathit{AP}$
is\\ $\varphi::=~p~|~\neg\varphi~|~
\varphi\vee\varphi~|~\op{X}\varphi~|~\varphi \op{U} \varphi$ for $p \in
\mathit{AP}$.
\end{definition}
\vspace{-.5em}

For $\Sigma = 2^\mathit{AP}$, a computation $u=u_0u_1..\in
\Sigma^\omega$ is a sequence where $u_i$ is the set of atomic propositions that hold at
the $i$-th position. For position $i$ we use $u,i\models \varphi$ to
denote that $\varphi$ holds at position $i$, inductively defined as: 
\vspace{-.2em}
\begin{pil}
  \item $u,i\models p$ iff $p \in u_i$;
  \item $u,i\models \neg \varphi$ iff $u, i \not\models \varphi$;
  \item $u,i\models \varphi_1 \vee \varphi_2$ iff
  $u,i\models \varphi_1$ or $u,i\models \varphi_2$;
  \item $u,i\models \op{X}\varphi$ iff $u, i{+}1\models \varphi$;
  \item $u,i\models \varphi_1\op{U}\varphi_2$ iff $\exists k\geq i{:~}
  u,k\models\varphi_2$ and
   $\forall j, i\leq j < k{:~} u,j\models\varphi_1.$
\end{pil}

We denote $u,0\models \varphi$ by $u \models \varphi$. We use additional
LTL operators \op{F} (finally) where $\op{F}\varphi :=
\true~\op{U}~\varphi$ and \op{G} (globally), where $\op{G}\varphi :=
\neg\op{F}\neg\varphi$. 

Finally, we also use PastLTL operators \op{Y} (previously, dual of
\op{X}) and \op{S} (since, dual of \op{U}):
\begin{pil}
  \item $u,i\models \op{Y}\varphi$ iff $i> 0 \wedge u,(i-1) \models \varphi$;
  \item $u,i\models \varphi_1 \op{S} \varphi_2$ iff $\exists k \leq i{:~}
  u,k\models\varphi_2$ and
   $\forall j, k < j \leq i{:~} u,j\models\varphi_1.$
\end{pil}

We use additional PastLTL operators \op{O} (once), where $\op{O}\varphi :=
\true~\op{S}~\varphi$, and \op{H} (historically), where $\op{H}\varphi :=
\neg \op{O} \neg \varphi$.

LTL formulas can be used as specifications of reactive systems where
atomic propositions are interpreted as environment (input) and
system (output) variables. An assignment to all variables is called
a state. 

A strategy for an LTL specification $\varphi$ prescribes the outputs of
a system that from its winning states for all environment choices lead
to computations that satisfy $\varphi$. A specification $\varphi$ is
called realizable if a strategy exists such that for all initial
environment choices the initial states are winning states.
The goal of LTL synthesis is, given an LTL specification, to find a
strategy that realizes it, if one exists.

\subsection{Generalized Reactivity of Rank 1 (GR(1))}

GR(1) is a fragment of LTL, which has an efficient symbolic synthesis
algorithm~\cite{BJP+12,PitermanPS06} and whose expressive power covers
most of the well-known LTL specification patterns of Dwyer et
al.~\cite{DAC99,MaozR15}.
GR(1) specifications include assumptions and guarantees about what needs
to hold on initial states, on all states (safety), and infinitely often
on every run (justice). A GR(1) synthesis problem consists of the
following elements~\cite{BJP+12}:
\begin{pil}
  \item $\mathcal{X}$ input variables controlled by the environment;
  \item $\mathcal{Y}$ output variables controlled by the system;
  \item $\theta^e$ assertion over $\mathcal{X}$ characterizing initial
  environment states;
  \item $\theta^s$ assertion over $\mathcal{X} \cup \mathcal{Y}$
  characterizing initial system states;
  \item $\rho^e(\mathcal{X}, \mathcal{Y}, \mathcal{X}')$ transition
  relation of the environment, where $\mathcal{X}'$ denotes a primed 
  copy of variables $\mathcal{X}$, i.e., given a current state $\rho^e$ restricts the next input;  
  \item $\rho^s(\mathcal{X}, \mathcal{Y}, \mathcal{X}',
  \mathcal{Y}')$ transition relation of the system, where $\mathcal{X}'$ 
  and $\mathcal{X}'$ denote primed copies of variables $\mathcal{X}$ 
  and $\mathcal{Y}$, i.e., given a current state and next input $\rho^s$ restricts the next output;  
  \item $J^e_{i \in 1..n}$ justice goals of the environment;
  \item $J^s_{j \in 1..m}$ justice goals of the system.
\end{pil}

%

A GR(1) synthesis problem is (strictly\footnote{For a comparison of
strict realizability and implication realizability
see~\cite{KleinP10,BJP+12}.}) realizable iff the following LTL formula
is realizable:
$$ \varphi^{sr} = (\theta^e \rightarrow \theta^s) \wedge (\theta^e
\rightarrow \op{G}((\op{H}\rho^e)\rightarrow \rho^s)) \wedge (\theta^e
\wedge \op{G}\rho^e \rightarrow (\bigwedge_{i \in 1..n} \op{GF} J_i^e
\rightarrow \bigwedge_{j \in 1..m} \op{GF} J_j^s)).
$$

Specifications for GR(1) synthesis have to be expressible in the above
structure and thus do not cover the complete LTL. Efficient symbolic
algorithms for GR(1) realizability checking and controller synthesis
have been presented in~\cite{BJP+12,PitermanPS06}.

GR(1) synthesis has been used and extended in different contexts and for
different application domains, including
robotics~\cite{Kress-GazitFP09,ManiatopoulosSP16}, scenario-based
specifications~\cite{MaozS12}, aspect languages~\cite{MaozS11}, event-based
behavior models~\cite{DIppolitoBPU13}, hybrid systems~\cite{FilippidisDLOM16},
and device drivers~\cite{WalkerR14}, to name a few.

\subsection{Extended Notation for Context-Free Grammars}
\label{sec:grammarNotation}

The Spectra language is defined by a context-free grammar. We now
describe the grammar notation that we use throughout this document. 
An excerpt of the Spectra grammar shown in Fig.~\ref{gra:example}.

Production rules are enclosed in brackets and printed bold. As an
example, the production rule for symbol \prodRule{specElem} is defined
in Fig.~\ref{gra:example}, l.~5 and referenced in l.~3 inside the
definition of the production rule for symbol \prodRule{spec}. We use
common multiplicity symbols for defining the number of repetitions of elements, where
$*$ denotes zero or more repetitions, $+$ denotes at least one
repetition, and $?$ denotes zero or one repetitions, i.e., the element
is optional. As an example, the
declaration of specification elements (\prodRule{specElem}, l.~3) is
repeated at least once. Terminals of the grammar, that are also keywords
of Spectra are printed in blue, e.g., the keywords \key{spec}, \key{sys}, and
\key{env}. Other terminals are printed in grey, e.g., the semicolon
(\texttt{\color{grey}{;}}, l.~7) after a variable declaration.
 
\begin{figure}\hspace*{\fill}\fbox{Grammar}\vspace{-0.85pt}
\begin{lstlisting}[language=grammar,mathescape]
<|spec|> ::= 
  $\key{spec}$ <|name|>
  ($elems$+=<|specElem|>)$^+$

<|specElem|> ::= <|varDec|> | <|assumption|> | <|guarantee|>

<|varDec|> ::= ($\key{sys}$ | $\key{env}$) <type> <name> $\synt{;}$

<|monitor|> extends <|specElem|> ::= ...
... 
\end{lstlisting}
\caption{Excerpt of the Spectra grammar to demonstrate our extended
notation for context-free grammars}
\label{gra:example}
\end{figure}

\paragraph{Grammar extensions} For readability and modular language
definition, we introduce two extensions to the grammar notation. First,
we introduce internal names for produced symbols, e.g., the name $elems$
for the list of \prodRule{specElem} in Fig.~\ref{gra:example}, l.~3. We
refer to these internal names when we describe well-formedness rules of
Spectra documents. Second, we introduce the grammar-level keyword
\textbf{\texttt{\color{purple}{extends}}} to denote the extension of
production rules. As an example, the production \prodRule{monitor}
extends \prodRule{specElem} in Fig.~\ref{gra:example}, l.~9. As a result
of the extension, the symbol \prodRule{monitor} can replace any
occurrence of the symbol \prodRule{specElem}. Note the difference
between \prodRule{varDec} and \prodRule{monitor} which are both
alternatives for \prodRule{specElem}. The difference is that production
rule \prodRule{varDec} was included in the definition of
\prodRule{specElem} while the production rule \prodRule{monitor} was
added later. We use this extension mechanism when we add new language
features.
Finally, we introduce the grammar-level keyword
\textbf{\texttt{\color{purple}{replaces}}} to denote that a new
production rule replaces an existing production rule.
As an example, the production \prodRule{specWithImports} replaces
\prodRule{spec} in Sect.~\ref{sec:imports}, Fig.~\ref{gra:import}, l.~1.
As a result of the replacement, the symbol \prodRule{specWithImports}
can replace any occurrence of the symbol \prodRule{spec}.

\paragraph{Special productions} We make use of three standard, primitive
productions that we do not define in any production rule. These are
\prodRule{file} for file names, \prodRule{name} for names of elements,
and \prodRule{int} for integers.

\paragraph{Simplification} The parts of the Spectra grammar we show in
Sect.~\ref{sec:kernel} and Sect.~\ref{sec:langElements} present a simplified
version of the grammar we use in the implementation of Spectra tools
described in Sect.~\ref{sec:tools}. For the purpose of presentation we
have simplified the production rules. As an example, it is common to
encode operator precedence directly into production rules to simplify
parsing of specifications. This implementation related formulation
blows up the grammar of expressions as every operator might require a
separate production rule. We present a simplified version of the grammar
and define operator precedence when we describe each language element.

\section{Example}
\label{sec:example}

We will introduce Spectra on the example of a specification for a Lego
forklift robot shown on the left side of Fig.~\ref{fig:combined}. The
forklift is an actual Lego robot\footnote{Robot based on these building
instructions: \url{http://www.nxtprograms.com/NXT2/forklift/steps.html}}
we have constructed and experimented with in our lab. We first reported
on this case study in~\cite{MaozR15synt}.

\subsection{Forklift Overview and Architecture}

\begin{figure}
  \centering
  \includegraphics[width=\textwidth]{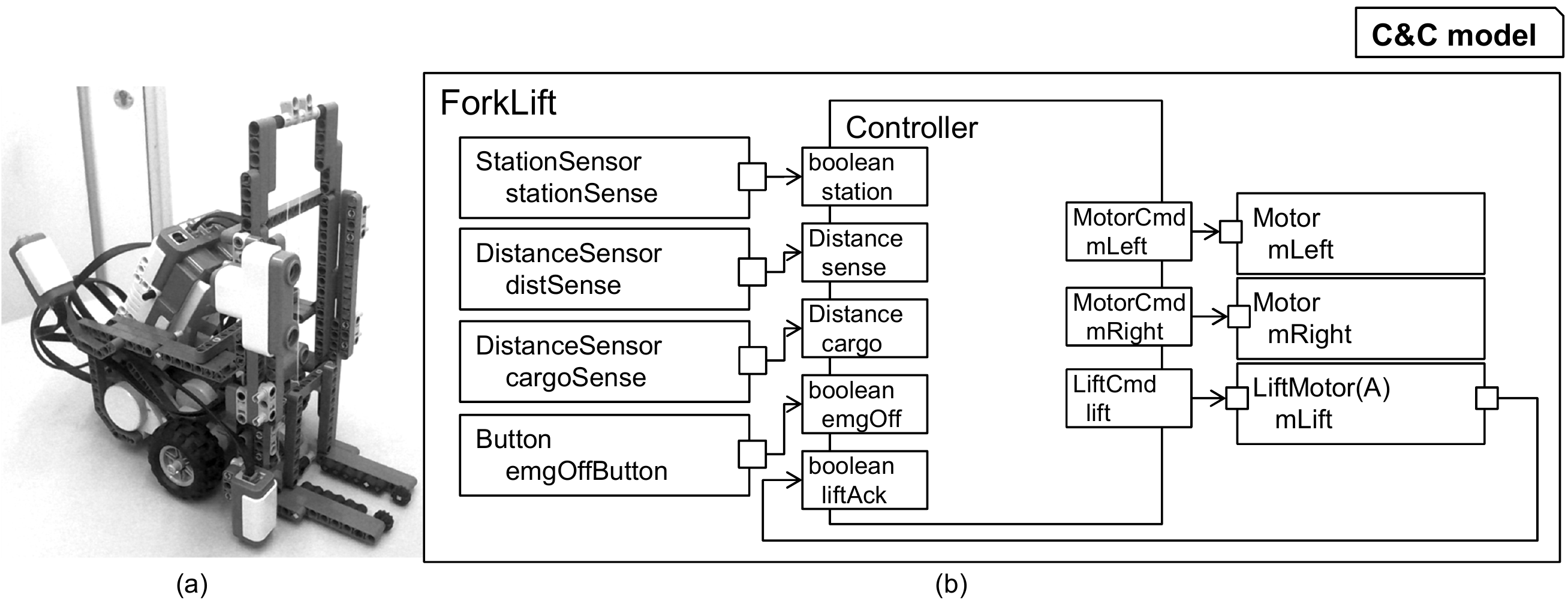}
  \caption{(a) The Lego forklift robot with four sensors and
  three actuators. (b) The component and connector model of the software architecture of the robot
  with wrappers for sensors and actuators. Data types definitions are shown in Fig.~\ref{fig:enums}. The main component
  \texttt{Controller} is to be specified in Spectra}
  \label{fig:combined}
\end{figure}

The forklift shown in Fig.~\ref{fig:combined} has a sensor to
determine whether it is at a station, two distance sensors to detect
obstacles and cargo, and an emergency button to stop it.
It has two motors to turn the left and right wheels and one motor to
lift the fork. Consider an initial set of informal requirements for the
behavior of the forklift:

\begin{enumerate}
  \item Do not run into obstacles.
  \item Only pick up or drop cargo at stations.
  \item Always keep on delivering cargo.
  \item Never drop cargo at the station where it was picked up.
  \item Stop moving if emergency off switch is pressed.
\end{enumerate} 

\begin{figure}
  \centering
  \includegraphics[width=.4\textwidth]{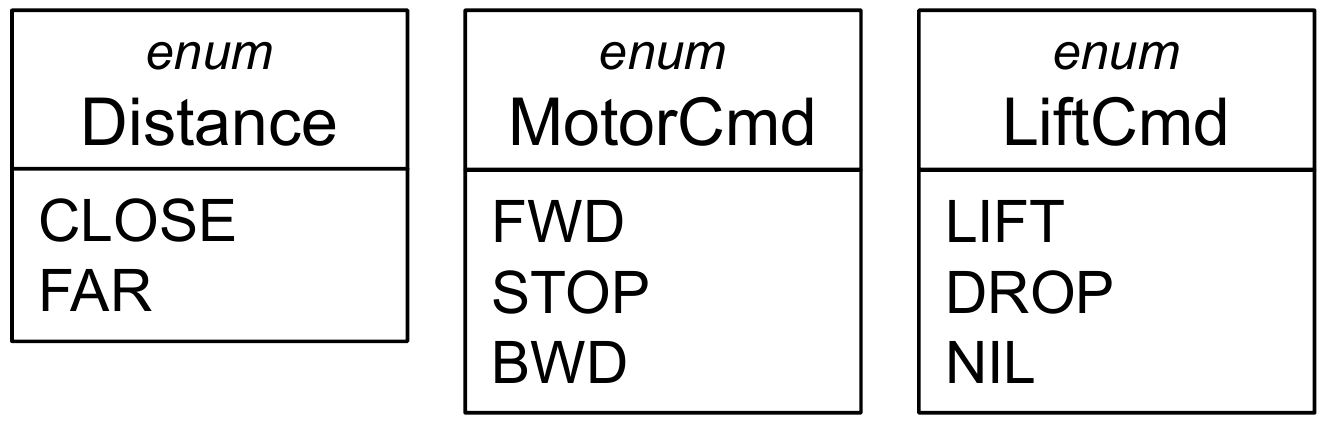}
  \caption{Data type definitions for ports of component
  \texttt{Controller} shown in Fig.~\ref{fig:combined}}
  \label{fig:enums}
\end{figure}

The logical software architecture of the forklift is depicted as a
component and connector model in Fig.~\ref{fig:combined}~(b). The
components on the left side are hardware wrappers that read sensor
values and publish them as messages on their output ports. The output
ports of these sensor components are connected to input ports of component
\texttt{Controller}. The output ports of component \texttt{Controller}
are connected to three components on the right that receive commands and
encapsulate access to the motors of the forklift. The datatypes of input
and output ports as well as their names are written on the ports of
component \texttt{Controller}. Datatypes other than \texttt{boolean} are
defined as enumerations in Fig.~\ref{fig:enums}.

The execution of the software components of the robot is performed in a
cycle: read sensor data, execute controller, perform actions. All
components are executed without any delay. This execution variant uses a
technique inspired by Raman et al.~\cite{RamanPK13} to synthesize a
reactive controller for continuous control. The setting requires the
synthesized controller to become aware whether actions have completed or
not. In our case driving actions get feedback from distance sensors and
a feedback signal from the lift motor acknowledges completion of the
motor's actions (see Fig.~\ref{fig:combined}).

\subsection{Forklift Specification in Spectra}

A specification for the reactive behavior of the forklift controller is
shown in Lst.~\ref{lst:vars} and Lst.~\ref{lst:spec}. This specification
is a subset of the complete specification
from~\cite{MaozR15synt}.\footnote{We have slightly adapted the
specification to use new language elements of Spectra, which where not
available yet for~\cite{MaozR15synt}.}

\begin{figure}\hspace*{\fill}\fbox{Spectra}\vspace{1.2pt}
\begin{lstlisting}[language=Spectra,label=lst:vars,
caption={Definition of data types, declaration of environment and 
system controlled variables, and definition of definitions}]
env Distance cargo;
env Distance sense;
env boolean station;
env boolean emgOff;
env boolean liftAck;

sys MotorCmd mLeft;
sys MotorCmd mRight;
sys LiftCmd lift;

type Distance = {FAR, CLOSE}; 
type MotorCmd = {FWD, STOP, BWD};
type LiftCmd = {LIFT, DROP, NIL};

define
  backing := mLeft = BWD & mRight = BWD;
  stopping := mLeft = STOP & mRight = STOP;
  turning := mLeft != mRight;
  forwarding := mLeft = FWD & mRight = FWD;
  dropping := lift = DROP;
  lifting := lift = LIFT;
  lowObstacle := (cargo = CLOSE & !station);
  deliverUnlessBlocked := (dropping | emgOff | lowObstacle);
\end{lstlisting}
\end{figure}

Lst.~\ref{lst:vars} shows the declaration of environment controlled
input variables (ll.~1-5) and system controlled output variables
(ll.~7-9). The names of environment variables in Lst.~\ref{lst:vars}
correspond to the names of input ports and the names of system variables
correspond to the names of output ports of component \texttt{Controller}
from Fig.~\ref{fig:combined}. Similarly, the enumeration types in
Lst.~\ref{lst:vars}, ll.~11-13 correspond to the enumeration types shown
in Fig.~\ref{fig:enums}. Finally, the remainder of Lst.~\ref{lst:vars}
introduces some useful defines, e.g., \texttt{backing} is defined as an
abbreviation of the expression \texttt{mLeft = BWD \& mRight = BWD},
which states that both motors receive the \texttt{BWD} command.

\begin{figure}\hspace*{\fill}\fbox{Spectra}\vspace{1.2pt}
\begin{lstlisting}[language=Spectra,label=lst:spec,
caption={Selection of assumptions, a monitor, and guarantees of the forklift specification from~\cite{MaozR15synt}}]
// station does not change when stopping
asm stationsDontMove:
  trans (stopping -> station = next(station));  

/** monitor waitingForLifting is true when we have lifted or dropped and
    not received a liftAck */
monitor boolean waitingForLifting {
  ini !waitingForLifting;
  trans ((lift=LIFT | lift=DROP) -> next(waitingForLifting));
  trans (liftAck -> (next(!waitingForLifting) | lift=LIFT | lift=DROP));
  trans (!(lift=LIFT | lift=DROP | liftAck) -> 
        waitingForLifting = next(waitingForLifting));
}

// assumptions on the input liftAck
asm ini !liftAck;
asm trans (next(liftAck) -> (waitingForLifting & !liftAck));
asm pRespondsToS(waitingForLifting, liftAck);

// (1) Do not run into obstacles 
gar dontHitObstacles:
  trans ((sense = CLOSE | lowObstacle) -> next(!forwarding));
  
// (2) Only pick up or drop cargo at stations.
gar liftDropAtStationOnly:
  alw ((lifting | dropping) -> atStation);

// (3) Always keep on delivering cargo.  
gar keepDelivering:
  alwEv deliverUnlessBlocked;  

// (4) Never drop cargo at the station where it was picked up.
gar leaveStationForDelivery: 
  pBecomesTrue_betweenQandR(!station, lifting, dropping);

// (5) Stop moving if emergency off is pressed.  
gar emergencyOff:
  alw (emgOff -> (stopping & lift=NIL));
  
/// monitor loaded is true iff we have cargo loaded and acknowledged
monitor boolean loaded {
  ini !loaded;  
  trans (liftAck & !loaded -> next(loaded));
  trans (liftAck & loaded -> next(!loaded));
  trans (!liftAck ->  loaded = next(loaded));
}

// restricting lifting action based on monitor loaded
gar restrictLifting: 
  alw ((loaded -> !lifting) & (!loaded -> !dropping));
\end{lstlisting}
\end{figure}

Lst.~\ref{lst:spec} shows a selection of assumptions, a monitor, and
guarantees of the forklift specification. As an example, the first
assumption named \texttt{stationsDontMove}, expresses that on all
transitions between states (operator \op{trans}) if the motors are
stopping then the sensor input for detecting a station does not change
in the next state (ll.~2-3). The monitor \texttt{waitingForLifting}
keeps track whether the forklift has recently issued a lifting command
\texttt{LIFT} or \texttt{DROP} and has not yet received an
acknowledgment through input \texttt{liftAck}. The monitor is referenced
inside a response pattern assumption in Lst.~\ref{lst:spec}, l.~18. This
assumption expresses that \texttt{waitingForLifting} is always followed
by a an acknowledgment \texttt{liftAck} after a finite time\footnote{In
LTL this would be expressed as \texttt{\op{G} (waitingForLifting ->
\op{F} liftAck)}. Note however, that this formula is syntactically not
part of the GR(1) fragment.}.

Next, Lst.~\ref{lst:spec} contains guarantees expressing the five
high-level requirements for the controller of the forklift. As an
example, the first requirement to not hit obstacles is formulated as the
guarantee \texttt{dontHitObstacles} in Lst.~\ref{lst:spec}, ll.~21-22.
It states that all transitions from sensing an obstacle or a low
obstacle are required to go to states where the forklift does not go
forward. The third requirement to always keep on delivering cargo is
formulated in Lst.~\ref{lst:spec}, ll.~29-30. The guarantee states that
always eventually (keyword \key{alwEv}) the forklift drops its cargo
unless it is blocked by obstacles (note that blocking the forklift
infinitely many times could prevent it from delivering cargo and we thus
had to weaken requirement (3)).

Finally, the specification in Lst.~\ref{lst:spec} shows the monitor
\texttt{loaded} and the guarantee \texttt{restrictLifting}. These were
added during experimenting with synthesized controllers. What happened
was that the synthesized controller decided to lift cargo when it had
already lifted cargo. This physically destroyed the lifting mechanism of
the robot. We realized that additional restrictions are necessary and
the guarantee \texttt{restrictLifting} thus encodes that the forklift
should never lift cargo when it already has cargo loaded and it should
never drop cargo when it does not have cargo loaded.

\section{Spectra Kernel}
\label{sec:kernel}

Spectra is based on a small kernel language for writing specifications.
The Spectra kernel contains only the essential elements to formulate
GR(1) synthesis problems.
Specifically, these language elements are Boolean environment variable
declarations, Boolean system variable declarations, assumptions, and
guarantees. We now introduce the syntax of the Spectra kernel and then
give the semantics of specifications as GR(1) synthesis problems.

\subsection{Syntax}

A grammar of the Spectra kernel is shown in Fig.~\ref{gra:kernel}. Every
valid specification can be produced from the production rule
\prodRule{spec} of the kernel grammar (and its extensions we present in
Sect.~\ref{sec:langElements}).

\begin{figure}\hspace*{\fill}\fbox{Grammar}\vspace{-0.85pt}
\begin{lstlisting}[language=grammar,mathescape]
<|spec|> ::= 
  $\key{spec}$ <|name|>
  ($elems$+=<|specElem|>)$^+$

<|specElem|> ::= <|varDec|> | <|assumption|> | <|guarantee|> 

<|varDec|> ::= ($\key{sys}$ | $\key{env}$) <type> <name> $\synt{;}$

<|type|>  ::= $\key{boolean}$
  
<|assumption|> ::= $\key{asm}$ (<|name|> $\synt{:}$)$^?$ 
  <|tempConstraint|> $\synt{;}$

<|guarantee|> ::= $\key{gar}$ (<|name|> $\synt{:}$)$^?$
  <|tempConstraint|> $\synt{;}$
  
<|tempConstraint|> ::=
  ($ini$     = $\key{ini}$ <|exp|>) |
  ($safety$  = $\key{trans}$ <|exp|>) | 
  ($justice$ = $\key{alwEv}$ <|exp|>) 
   
<|exp|> ::= $\synt{(}$ <|exp|> $\synt{)}$ |
  $left$=<|exp|> $op$=<|binaryOp|> $right$=<|exp|> |
  <|unaryOp|> <|exp|>
  <|primExp|>
  
<|unaryOp|> ::= $\synt{!}$ | $\op{next}$

<|binaryOp|> ::= $\synt{\&}$ | $\synt{|}$ | $\synt{->}$ | $\synt{=}$ | $\synt{<->}$

<|primExp|> ::= $\key{true}$ | $\key{false}$ | $varRef$=<|name|> 
\end{lstlisting}
\caption{Grammar of the Spectra kernel}
\label{gra:kernel}
\end{figure}

Every specification starts with the keyword \key{spec} and a
name. The name of a specification is only informative and has no technical meaning. A
specification contains one or more specification elements
\prodRule{specElem}. Specification elements of the Spectra kernel are
variable declarations (Fig.~\ref{gra:kernel}, l.~7), assumptions
(Fig.~\ref{gra:kernel}, l.~11), and guarantees (Fig.~\ref{gra:kernel},
l.~16).

Variable declarations introduce variables controlled by the environment
(keyword \key{env}) or by the system (keyword \key{sys}). Variables have
a type and a name. The only type available in the kernel is
\key{boolean} (Fig.~\ref{gra:kernel}, l.~9).

Assumptions have an optional name and can be of three kinds. They are
either initial assumptions (keyword \key{ini}), safety assumptions
(keyword \key{trans}), or justice assumptions (keyword
\key{alwEv}).\footnote{Other GR(1)-based specification
languages use the LTL operator \key{G} for \key{trans} and \key{GF} for
\key{alwEv}. This leads to confusion as their semantics
usually does not match the LTL semantics.} All assumption kinds are formulated by
expressions \prodRule{exp} (Fig.~\ref{gra:kernel}, l.~21).
Expressions follow standard definitions of propositional expressions.
The unary operators of expressions are negation and \op{next}, the
binary operators are standard Boolean operators and equality, and the
primary expressions are \key{true}, \key{false}, and references to
variables (Fig.~\ref{gra:kernel}, l.~30). The precedence of operators in
expressions from strongest binding to weakest is \synt{!}, \op{next},
\synt{=}, \synt{\&}, \synt{|}, \synt{<->}, \synt{->}. As usual, binary
operators associate from left to right.

Guarantees are defined analogously to assumptions with the exceptions of
the keyword \key{gar} instead of \key{asm} and differences in
well-formedness rules.

\paragraph{Well-formedness rules}
The names of variables, assumptions, and guarantees have to be unique.
An initial assumption cannot reference system variables. A safety
assumption cannot nest references to system variables in the scope of
the \op{next} operator. A \op{next} operator cannot be (transitively)
nested inside a \op{next} operator.

\paragraph{Scopes} 
The Spectra kernel contains three kinds of named elements: variables,
assumptions, and guarantees. The scope of variable names is global,
i.e., the names of variables have to be unique and variables can be
referenced from anywhere in the specification. The scope of names of
assumptions and guarantees is also global but they cannot be referenced
from anywhere inside the specification.

\subsection{Semantics}

The semantic domain of the Spectra kernel language are GR(1) synthesis
problems. We define the semantics of the Spectra kernel by
translating every well-formed specification to a GR(1) synthesis problem. 

Given a Spectra specification $m$ its semantics is defined by the following GR(1) synthesis problem:
\begin{pil}
  \item $\mathcal{X}$ is the set of all variables declared in $m$ with the keyword \key{env};
  \item $\mathcal{Y}$ is the set of all variables declared in $m$ with the keyword \key{sys};
  \item $\theta^e$ is the conjunction of the semantics of all initial assumptions;
  \item $\theta^s$ is the conjunction of the semantics of all initial guarantees;
  \item $\rho^e(\mathcal{X} \cup \mathcal{Y}, \mathcal{X})$ is the conjunction of the semantics of all safety assumptions;  
  \item $\rho^s(\mathcal{X} \cup \mathcal{Y}, \mathcal{X} \cup
  \mathcal{Y})$  is the conjunction of the semantics of all safety guarantees;  
  \item $J^e_{i \in 1..n}$ is the set of the semantics of all justice assumptions;
  \item $J^s_{j \in 1..m}$ is the set of the semantics of all justice guarantees.
\end{pil} 

To obtain the semantics of assumptions and guarantees we translate their
expression \prodRule{exp} to assertions in the GR(1) synthesis problem.
This translation is straight forward on the structures of
\prodRule{exp}. The symbols of Boolean operators of \prodRule{unaryOp}
and \prodRule{binaryOp} are translated to their counterparts in
assertions. The keywords \key{false} and \key{true} of \prodRule{primExp}
are translated to a contradiction and a tautology, respectively. A
variable reference of \prodRule{primExp} is translated to an assertion
over Boolean variable from $\mathcal{X}$ and $\mathcal{Y}$ with the
referenced name. Finally, the unary operator \op{next} substitutes all
variables from $\mathcal{X}$ and $\mathcal{Y}$ in its scope with
corresponding primed variables from $\mathcal{X}'$ and $\mathcal{Y}'$.

\section{Spectra Language Elements}
\label{sec:langElements}

We now present the Spectra language elements beyond the kernel language. For
each language element we discuss the motivation to have it in the language,
present its syntax as a grammar that extends the Spectra kernel, list
well-formedness rules of the extended language, discuss the scope of newly
introduced named elements, and finally present the semantics of new elements.

Note that we define the semantics of Spectra language elements by a
translation to Spectra without the new language element. A direct
translation to the Spectra kernel would be possible but often more
complicated because all combinations of language elements would have to
be considered in the translation, e.g., references to defines can appear
in parameters of predicates.
Instead, we present modular translations that allow for the combination
of any subset of Spectra language elements. The application of all
semantics defining translations will result in a specification in the
Spectra kernel language. This design also ensures that extensions of the
Spectra kernel beyond GR(1) will directly support the new Spectra
language elements.~\footnote{Examples of extensions are GR(k)
specifications~\cite{PitermanP06} or cooperative
synthesis~\cite{BloemEK15}.}


\subsection{Enumerations and bounded Integers}
\label{sec:enums}

\subsubsection{Motivation}

The Spectra kernel only contains Boolean variables. However, in many
cases it is more convenient to use enumerations of values to define the
type of a multi-valued variable. As an example, we define the enumeration type
\texttt{MotorCmd} in Lst.~\ref{lst:vars}, l.~12 with values \texttt{FWD},
\texttt{STOP}, and \texttt{BWD}. In expressions, these values can be
used and compared to each other and to variables of the same type,
e.g., see Lst.~\ref{lst:vars}, ll.~16 for the definition of
\texttt{backing} where both motor variables have the value \texttt{BWD}.

Similarly, some values are best modeled using integers. Spectra supports
a bounded integer type with lower and upper bounds. An example appears
in Lst.~\ref{lst:int} where the variable \texttt{speed} is of integer
type with values from 0 to 50. In expressions, integer variables can be
compared with primitive values, e.g., see Lst.~\ref{lst:int}, l.~3 where
the value of variable \texttt{speed} is restricted to at most 6 when
close to cargo.
Bounded integers can also be compared to arithmetic expressions and
other integer variables even when their bounds are different.

\begin{figure}\hspace*{\fill}\fbox{Spectra}\vspace{1.2pt}
\begin{lstlisting}[language=Spectra,label=lst:int,
caption={An example of an integer variable used in a guarantee.}]
sys Int(0..50) speed;

gar alw cargo=CLOSE -> (speed <= 6);
\end{lstlisting}
\end{figure}

\subsubsection{Syntax} 
The syntax of enumerations and bounded Integers is shown in
Fig.~\ref{gra:enumsAndInt}.
Enumerations are defined by production rule \prodRule{enumType} and consist of a
list of names enclosed in curly brackets \synt{\{}, \synt{\}}. Names of
enumeration values can be referenced as primary expressions
(Fig.~\ref{gra:enumsAndInt}, ll.~5-6), e.g., they can also be used in
expressions.

\begin{figure}\hspace*{\fill}\fbox{Grammar}\vspace{-0.85pt}
\begin{lstlisting}[language=grammar,mathescape]
// enumerations have one or more named values
<|enumType|> extends <|type|> ::= 
  $\synt{\{}$ $vals$=<|name|> ($\synt{,}$ $vals$+=<|name|>)$^*$ $\synt{\}}$

<|enumValRef|> extends <|primExp|> ::= 
  $enumValRef$=<|name|>

// integers are bounded and support arithmetic operations
<|intType|> extends <|type|> ::= 
  $\synt{Int(}$$lower$=<|int|>$\synt{..}$$upper$=<|int|>$\synt{)}$

<|intBinaryOp|> extends <|binaryOp|> ::=
  $\synt{+}$ | $\synt{-}$ | $\synt{*}$ | $\synt{/}$ | $\synt{mod}$ | $\synt{<}$ | $\synt{>}$ | $\synt{<=}$ | $\synt{>=}$
  
<|intUnaryOp|> extends <|unaryOp|> ::=  
  $\synt{-}$
  
<|intVal|> extends <|primExp|> ::= 
  <|int|>
  
\end{lstlisting}
\caption{Syntax of enumeration types and bounded integers}
\label{gra:enumsAndInt}
\end{figure}

Bounded Integer types in Spectra start with \synt{Int} followed by lower
and upper bounds in parenthesis (Fig.~\ref{gra:enumsAndInt}, ll.~9-10).
We also add arithmetic operators for addition, subtraction,
multiplication, division, and modulo to Spectra, together with
inequalities in Fig.~\ref{gra:enumsAndInt}, l.~12-13. The inequalities
bind with the same priority as equality and the arithmetic operators
bind stronger than all previously introduced operators. Finally, we add
arithmetic negation as a unary operator and integer constants as primary
expressions in Fig.~\ref{gra:enumsAndInt}, ll.~15-19.

\paragraph{Well-formedness rules}
The names of enumeration values must be unique and different from variable
names.
Equality is the only comparison allowed for enumeration values and
variables of enumeration type and Boolean type.
The upper bound $upper$ of a bounded integer must be strictly greater
than its lower bound $lower$.
The new operators \prodRule{intBinaryOp} and \prodRule{intUnaryOp} can
only be used in arithmetic expressions constructed from these operators,
primitive values \prodRule{intVal}, and references to Integer variables
\prodRule{varRef}.


\subsubsection{Semantics}
We define the semantics of enumerations and bounded Integers by a
translation into Spectra without enumerations and bounded Integers. We
translate every variable of an enumeration type with $n=|vals|$ values to
$\ceil{\log_2(n)}$ Boolean variables. To every value in $vals$ we assign
one combination of values of the new variables. We add initial and
safety constraints to the specification that the new variables represent one of the defined
values (important for $2^{\ceil{\log_2(n)}} > n$). We translate every
equality expression involving enumerations to equivalent expressions
over the Boolean variables that encode them.

Similarly, we translate every bounded integer variable to
${\ceil{\log_2(upper-lower)}}$ Boolean variables. For the translation of
bounded arithmetic expressions, we employ an encoding inspired
by~\cite{BartzisB06}.

\subsection{Defines and type definitions}
\label{sec:defines}

\subsubsection{Motivation}

Specifications often contain sub-expressions that reoccur many times. In
the example of the forklift, many guarantees and assumptions refer to
stopping both motors, formally \texttt{mLeft = STOP \& mRight = STOP}.
Once these reoccurring expressions get longer and more complicated, the
overall specification becomes harder to read and difficult to maintain,
e.g., when updating a condition. To support more concise specifications
and to simplify the maintenance of reoccurring sub-expressions, Spectra
introduces defines. In the forklift example a define introduces
the name \texttt{stopping} for the expression \texttt{mLeft = STOP \&
mRight = STOP}. Assumptions and guarantees then use \texttt{stopping} as
a reference to the define.

Similarly, several variables in a specification may be of the same type. For
example, in the forklift specification the left and right motors both have the
enumeration type \texttt{\{FWD, STOP, BWD\}}. Here, to support more concise
specifications and to simplify the maintenance of reoccurring types, Spectra introduces
type definitions. In the forklift example, the type \texttt{MotorCmd} is defined
by a type definition in Lst.~\ref{lst:vars}, l.~12 and referenced as the type of
system variables \texttt{mLeft} and \texttt{mRight} in ll.~7-8.

\subsubsection{Syntax}

The syntax of defines and type definitions is shown in
Fig.~\ref{gra:defines}. Defines are specification elements and thus
appear as top level elements inside specifications. Defines have a name
and an expression.
References to defines \prodRule{defRef} are primary expressions
(Fig.~\ref{gra:defines}, l.~4). 

\begin{figure}\hspace*{\fill}\fbox{Grammar}\vspace{-0.85pt}
\begin{lstlisting}[language=grammar,mathescape]
<|def|> extends <|specElem|> ::= 
  $\key{define}$ <|name|> $\synt{:=}$ <|exp|> $\synt{;}$

<|defRef|> extends <|primExp|> ::= 
  $defRef$=<|name|>

// type definitions introduce an alias for a type 
<|typeDef|> extends <|specElem|> ::= 
  $\key{type}$ <|name|> $\synt{=}$ <|type|> $\synt{;}$

// a reference to the alias can be used where types are expected
<|typeRef|> extends <|type|> ::= 
  $typeRef$=<|name|>

\end{lstlisting}
\caption{Syntax of defines and type definitions}
\label{gra:defines}
\end{figure}

Type definitions (Fig.~\ref{gra:defines}, l.~8) are also
specification elements and thus appear as top level elements inside
specifications. Type definitions have a name and a type. 
References to type definitions \prodRule{typeRef} are treated as types 
(Fig.~\ref{gra:defines}, l.~12).

\paragraph{Well-formedness rules}
The names of defines and type definitions must be unique names of
elements of the specification.
Defines with the operator \op{next} may not be used in expressions where
\op{next} is not allowed, e.g., defines with the operator \op{next} may
not be nested inside the operator \op{next}. Note that the expression of
a define is not required to evaluate to a Boolean value. However, the
semantics of a specification with defines and type definitions has to be
a well-formed specification.

\paragraph{Scopes} The names of defines and type definitions have the
same scope as the names of environment and system variables defined in
the Spectra kernel. The names of defines are visible to other defines. 

\subsubsection{Semantics}
We define the semantics of defines and type definitions by a translation
into Spectra without defines and type definitions. Every reference to a
define is replaced by the expression of the define. Every reference to a
type definition is replaced by the type of the type definition.

\subsection{State Invariants}
\label{sec:always}

\subsubsection{Motivation}

Some constraints should hold in all states, e.g., the guarantee that the
forklift lifts and drops cargo only at stations. These assumptions and
guarantees could be expressed as combinations of initial and safety
constraints, however we chose to introduce a separate keyword for state
invariants to make specifications easier to read and
write.\footnote{Note that the common keyword \key{G} used for
\key{trans} in other GR(1) specification languages is often confused for
defining state invariants (as the LTL semantics would suggest) leading
to problems. A discussion is outside the scope of this paper.}

\subsubsection{Syntax}

We show the syntax of state invariants in Fig.~\ref{gra:stateInv}.
State invariants are temporal constraints by extension of
\prodRule{tempConstraint} and can be used anywhere these constraints
appear, e.g., in assumptions and guarantees or in patterns (see
Sect.~\ref{sec:patterns}).

\begin{figure}\hspace*{\fill}\fbox{Grammar}\vspace{-0.85pt}
\begin{lstlisting}[language=grammar,mathescape]
// new temporal constraint for state invariants 
<|stateInvConstraint|> extends <|tempConstraint|> ::=
  $stateInv$  = $\key{alw}$ <|exp|> 
\end{lstlisting}
\caption{State invariants are an extension of temporal constraints}
\label{gra:stateInv}
\end{figure}

\paragraph{Well-formedness rules} The expression \prodRule{exp} of a
state invariant may not contain the operator \key{next}. A state
invariant of the environment, i.e., used in an assumption, may not
contain system variables.
 
\subsubsection{Semantics}
We define the semantics of state invariants by a translation of state
invariants to Spectra without state invariants. Every state invariant is
translated into an initial constraint with the same expression
\prodRule{exp} and a safety constraint (with keyword \op{trans}) with
the expression \prodRule{exp} nested inside the operator \key{next}.
Together the two constraints ensure that the invariants holds in the
initial state and in every successor state.

\subsection{PastLTL}
\label{sec:pastLTL}

\subsubsection{Motivation}

Assumptions and guarantees define constraints on reactive behavior by
relating the inputs and outputs of the current state to those of the
next state. However, in some situations the restriction to only current
and next inputs and outputs becomes limiting. As an example, the
forklift is not allowed to lift cargo when it has cargo lifted already.
Yet, in our example, there is no input indicating whether cargo is
currently loaded\footnote{The example introduces a monitor for this
purpose. Here we discuss an alternative based on PastLTL.}.
Interestingly, it is easy to determine from past observations whether
cargo is currently loaded: if in the previous state cargo has not been
dropped since it had been lifted in the past then cargo is currently
loaded. We can formalize the above using PastLTL operators supported by
Spectra:
$\texttt{\op{PREV} (lift!=DROP \op{SINCE} lift=LIFT)}$.

\subsubsection{Syntax}

We show the syntax of PastLTL expressions in Fig.~\ref{gra:PastLTL}.
Spectra supports three unary PastLTL operators and one binary PastLTL
operator. For all operators Spectra also provides a one letter
abbreviation (preceding the operator name in Fig.~\ref{gra:PastLTL}).

\begin{figure}\hspace*{\fill}\fbox{Grammar}\vspace{-0.85pt}
\begin{lstlisting}[language=grammar,mathescape]
// unary PastLTL operators
<|pLTLUnaryOp|> extends <|unaryOp|> ::=
  $\key{Y}$ | $\key{PREV}$ |
  $\key{H}$ | $\key{HISTORICALLY}$ |
  $\key{O}$ | $\key{ONCE}$  

// binary PastLTL operators
<|pLTLBinaryOp|> extends <|binaryOp|> ::=
  $\key{S}$ | $\key{SINCE}$
\end{lstlisting}
\caption{PastLTL operators introduced as extensions of unary and binary
operators}
\label{gra:PastLTL}
\end{figure}

\paragraph{Well-formedness rules}
All operands of PastLTL operators must evaluate to Boolean values (all
PastLTL expressions evaluate to Boolean values).

\subsubsection{Semantics}

We define the semantics of PastLTL by a translation of PastLTL to
Spectra without PastLTL. Recall from the preliminaries that the two
operators \op{PREV} and \op{SINCE} provide full expressiveness of
PastLTL; all other PastLTL operators can be defined in terms of these
two. We will show the translation for the operators \texttt{\op{PREV}}
and \texttt{\op{SINCE}}.

Given the PastLTL expression \texttt{\op{PREV} $\varphi$}, we create a
Boolean system variable \texttt{aux} with a fresh name. We then add the
initial guarantee \texttt{\op{ini} !aux} and the safety guarantee
\texttt{\op{trans} \op{next}(aux) <=> $\varphi$}. Technically, the
variable \texttt{aux} stores the previous valuation of $\varphi$.
Finally, we replace all occurrences of \texttt{\op{PREV} $\varphi$} with
a reference to the new variable \texttt{aux}.

Given a PastLTL expression \texttt{$\varphi$ \op{SINCE} $\psi$}, we
create a Boolean system variable \texttt{aux} with a fresh name. We then
add the initial guarantee \texttt{\op{ini} aux <=> $\psi$} and the
safety guarantee \texttt{\op{trans} \op{next}(aux) <=> (aux \&
\op{next}($\varphi$) | \op{next}($\psi$))}. The variable \texttt{aux} is
true if $\psi$ holds or if \texttt{aux} was true before and $\varphi$
holds, i.e., if $\varphi$ holds since $\psi$ was true in the past.
Finally, we replace all occurrences of \texttt{$\varphi$ \op{SINCE}
$\psi$} with a reference to the new variable \texttt{aux}.

\subsection{Predicates}
\label{sec:predicates}

\subsubsection{Motivation}

Sometimes repeating parts appearing in specifications differ in
sub-expressions. In these cases defines from Sect.~\ref{sec:defines}
cannot be used to avoid repetition. For these cases we introduce
predicates, which can instantiate sub-expressions with their parameters.

As an example, consider an extension of the forklift to load items in
the front and in the back, as shown in Lst.~\ref{lst:pred}. To know
whether an item is loaded we express that it is loaded in the front or
loaded in the back. To avoid repetition we can extract the expression
whether an item is loaded into a predicate with a parameter of the item
as shown in Lst.~\ref{lst:pred}, ll.~5-7. The predicate can then be
instantiated, e.g, in a guarantee as shown in l.~10, for different
items.

\begin{figure}\hspace*{\fill}\fbox{Spectra}\vspace{1.2pt}
\begin{lstlisting}[language=Spectra,label=lst:pred,
caption={An example of a predicate definition and predicate instances in a guarantee.}]
type Item = {Livestock, Chemicals};
sys Item loadFront;
sys Item loadBack;

predicate loaded(Item i) {
  loadFront = i | loadBack = i 
}

gar carrySafety:
  alw loaded(Livestock) -> !loaded(Chemicals);
\end{lstlisting}
\end{figure}

\subsubsection{Syntax}

We show the syntax of predicate definitions and predicate instantiations
in Fig.~\ref{gra:predicate}. The grammar starts with the syntax for
predicate definitions \prodRule{predicate}, which are specification
elements and thus appear as top level elements inside specifications.
Predicate definitions have a name followed by a list of typed parameters
\prodRule{typedParam}. The body of a predicate definition is an
expression.

Predicate instances \prodRule{predInst} (Fig.~\ref{gra:predicate}, ll.~11-12)
are primary expressions. Instances reference the name of a predicate and
provide expressions for each parameter of the predicate.

\begin{figure}\hspace*{\fill}\fbox{Grammar}\vspace{-0.85pt}
\begin{lstlisting}[language=grammar,mathescape]
// predicates have typed parameters and a propositional expression
<|predicate|> extends <|specElem|> ::= 
  $\key{predicate}$ <|name|> $\synt{(}$ $params$+=<|typedParam|> ($\synt{,}$ $params$+=<|typedParam|>)$^*$ $\synt{) \{}$
    $body$=<|exp|>
  $\synt{\}}$

// typed parameters have a type and a name
<|typedParam|> ::= <|type|> <|name|>

// predicates can be instantiated with expressions as parameters
<|predInst|> extends <|primExp|> ::= 
  <|name|> $\synt{(}$ $pVals$+=<|exp|> ($\synt{,}$ $pVals$+=<|exp|>)$^*$ $\synt{)}$
\end{lstlisting}
\caption{Syntax of predicates with names, parameters, and a predicate
body over parameters}
\label{gra:predicate}
\end{figure}

\paragraph{Well-formedness rules}
All parameters of predicate instances must be expressions that evaluate
to the type declared in the predicate definition. The body of the
predicate must be an expression that evaluates to Boolean. The body of a
predicate cannot (also not transitively, e.g., through predicate
instances or defines) contain an instance of the predicate itself.

\paragraph{Scopes}
The names of predicates have the same scope as the names of environment
and system variables defined in the Spectra kernel. As a result, the
names of predicates are visible to other predicates. The names of
parameters of the predicate are only visible inside the body of the
predicate.

\subsubsection{Semantics}
We define the semantics of predicate instantiations and predicates by a
translation into Spectra without predicates. Every predicate instance is
translated into a Boolean expression. Technically, we replace a
predicate instance by a copy of the body of the predicate. In the copy,
we replace references to predicate parameters by the expressions provided
for each parameter.

\subsection{Monitors}
\label{sec:monitors}

\subsubsection{Motivation}

Assumptions and guarantees define constraints on reactive behavior by
relating the inputs and outputs of the current state to those of the
next state. In some situations the restriction of referencing only
inputs and outputs becomes limiting. As an example, the forklift is not
allowed to lift cargo when it has cargo lifted already. However, in our
example, there is no input variable indicating whether cargo has been
lifted. As a solution, Spectra introduces monitors which come with a
definition of how their value is updated at every execution step. As an
example, the monitor \texttt{loaded} (Lst.~\ref{lst:spec}, l.~41-46) is
of type \texttt{boolean} and its value is updated on receiving
acknowledgments of lift actions. This monitor is used in a guarantee
(Lst.~\ref{lst:spec}, l.~50) to prevent lifting cargo when cargo is
already lifted and to prevent dropping cargo when no cargo was lifted.

\subsubsection{Syntax}
The syntax of monitors is shown in Fig.~\ref{gra:monitor}. Monitors are
specification elements and thus appear as top level elements inside
specifications. Monitors have a type and a name. The body of a monitor
consists of initial and safety constraints.

References to monitors \prodRule{monRef} are primary expressions
(Fig.~\ref{gra:monitor}, l.~8).

\begin{figure}\hspace*{\fill}\fbox{Grammar}\vspace{-0.85pt}
\begin{lstlisting}[language=grammar,mathescape]
// monitors have a type, a name, and constraints
<|monitor|> extends <|specElem|> ::= 
  $\key{monitor}$ <|type|> <|name|> $\synt{\{}$
    ($cons$+=<|tempConstraint|>  $\synt{;}$)$^+$     
  $\synt{\}}$

// references to monitors are primary expressions
<|monRef|> extends <|primExp|> ::= <|name|>
\end{lstlisting}
\caption{Syntax of monitors with types, names, and monitor constraints}
\label{gra:monitor}
\end{figure}

\paragraph{Well-formedness rules}
The names of monitors must be unique names of elements of the
specification.
The body of a monitor may not contain justice constraints.
The expression \prodRule{exp} of an initial constraint may not include
the unary operator \op{next}.
The semantics of the constraints in the body of the monitor must assign a
unique value to the monitor in any step and it must not restrict any
other variable (when seen as an automaton with the monitor variable as
states and all other variables as input, the defined automaton has to be
deterministic and complete).\footnote{This property is difficult to
check or ensure syntactically.}

\paragraph{Scopes} The names of monitors have the same scope as the
names of environment and system variables defined in the Spectra kernel.
Importantly, the name of a monitor is visible in the constraints of the
monitor.

\subsubsection{Semantics}
We define the semantics of monitors by a translation of monitors to Spectra
without monitors. For every monitor definition we create a system
variable with the name and type of the monitor. All monitor constraints
become guarantees. All references to the monitor become references to
the new system variable with the same name and type.

\subsection{Patterns}
\label{sec:patterns}

\subsubsection{Motivation}
Specifications of reactive systems require engineers to express complex
temporal relations between environment and system states. LTL provides
great expressiveness for temporal relations. However, some basic
relations lead to long formulas, which might be complicated to read and
challenging to write correctly. As an example, consider the guarantee
that the forklift should leave the station (\texttt{!atStation}) between
lifting cargo (\texttt{lifting}) and dropping cargo (\texttt{dropping}).
This guarantee can be expressed in LTL as follows\footnote{LTL formula
taken from the patterns catalog of Dwyer et al.~\cite{DAC99}. There
might be multiple ways to express an equivalent guarantee.}:
\begin{multline*}
\texttt{\op{G} (lifting \& !dropping ->}\\ 
\texttt{((!dropping \op{U} (!station \&
!dropping)) | \op{G} !dropping))}
\end{multline*}
Note that assertions on relevant states, e.g., \texttt{lifting} or
\texttt{dropping}, are repeated and that LTL operators are nested, e.g.,
the operator until (\op{U}) is nested inside the scope of the first
globally operator (\op{G}). 

Dwyer et al.~\cite{DAC99} have identified 55 LTL specification patterns
which are common in industrial specifications. They have suggested a
classification and natural language descriptions for the identified
patterns. The example above is one of their patterns and would be
transcribed as:
$$\texttt{!atStation \op{occurs between} lifting \op{and} dropping}$$

We have investigated patterns for GR(1) synthesis in~\cite{MaozR15}. We have
shown that 52 of these 55 LTL specification patterns can be supported as
assumptions and guarantees for GR(1) synthesis. In Spectra, the example
guarantee is formulated as a pattern definition (Spectra comes with a catalog of
pattern definitions for all 52 supported LTL specification patterns) and the
pattern instantiation shown in Lst.~\ref{lst:spec}, l.~18-19.

Note that Spectra patterns are not limited to the 52 patterns defined in
the catalog. Spectra provides engineers with means to define and
reference their own patterns. 

\subsubsection{Syntax}

The syntax of patterns is shown in Fig.~\ref{gra:pattern}. The grammar
starts with the syntax for pattern definitions \prodRule{pattern}, which
are specification elements and thus appear as top level elements inside
specifications. Pattern definitions have a name followed by a list of
parameter names. The body of a pattern definition can declare variables
local to the pattern and temporal constraints. A pattern may contain any
number of initial and safety constraints, but it must contain exactly
one justice constraint.

Pattern instances \prodRule{patInst} (Fig.~\ref{gra:pattern}, ll.~12-13)
are primary expressions. Instances reference the name of a pattern and
provide expressions for each parameter of the pattern.

\begin{figure}\hspace*{\fill}\fbox{Grammar}\vspace{-0.85pt}
\begin{lstlisting}[language=grammar,mathescape]
// pattern definitions have parameters and constraints
<|pattern|> extends <|specElem|> ::= 
  $\key{pattern}$ <|name|>$\synt{(}$$params$+=<|name|> ($\synt{,}$ $params$+=<|name|>)$^*$ $\synt{) \{}$
    $vars$+=<|patVar|>$^+$
    ($cons$+=<|tempConstraint|>  $\synt{;}$)$^+$ 
  $\synt{\}}$
  
<|patVar|> ::= 
  $\key{var}$ <|type|> <|name|>$\synt{;}$ 

//pattern instantiation with parameters is a primary expression
<|patInst|> extends <|primExp|> ::= 
  <|name|> $\synt{(}$$pVals$+=<|exp|> ($\synt{,}$ $pVals$+=<|exp|>)$^*$ $\synt{)}$
\end{lstlisting}
\caption{Syntax of patterns with names, parameters, auxiliary
variables, and pattern constraints over auxiliary variables and
parameters}
\label{gra:pattern}
\end{figure}

\paragraph{Well-formedness rules} 
The body of a pattern contains exactly one justice constraint.
The expression \prodRule{exp} of an initial or justice constraint may
not include the unary operator \op{next}. The expression \prodRule{exp}
of a safety constraint may only include references to pattern variables $vars$
in the scope of the operator \op{next}.
The semantics of the initial and the safety constraints must assign a
unique value to all pattern variables in any step and it must not
restrict any other variable (when seen as an automaton with the pattern
variables as states and all other variables as input, the defined
automaton has to be deterministic and complete).\footnote{This property
is difficult to check syntactically.}

The production \prodRule{patInst} may only appear exclusively as a
primary expression of assumptions or guarantees, i.e., it cannot
appear inside another expression and not in any other place in a
specification.

\paragraph{Scopes} The names of pattern parameters $params$ are only visible
inside the body of the pattern delimited by \synt{\{}, \synt{\}}.
The names of pattern variables $vars$ are only visible inside the body of the
pattern delimited by \synt{\{}, \synt{\}}. The names of patterns are visible
beyond specification files and thus can be imported, see~\ref{sec:imports}.

\subsubsection{Semantics}
We define the semantics of patterns by a translation of pattern
instances and patterns to Spectra without patterns. For every pattern
instance we create a copy of all pattern variables as system variables with a
fresh name and the same type. We instantiate pattern constraints by
replacing references to pattern variables with references to the fresh
names and by replacing references to variable names with copies of the
corresponding expression of the parameter values from $pVals$.

Instantiated initial pattern constraints become initial guarantees;
instantiated safety pattern constraints become safety guarantees; the
instantiated justice constraint becomes a justice assumption if the
pattern instance is contained inside an assumption; and the
instantiated justice constraint becomes a justice guarantee if the
pattern instance is contained inside a guarantee (see
also~\cite{MaozR15}).

\subsection{Imports}
\label{sec:imports}

\subsubsection{Motivation}

We have motivated many language features of Spectra based on reuse. In
all cases a reuse within a single specification makes sense. In some
cases, a reuse across specifications is also desirable. As an example,
patterns, e.g., from the catalog of Dwyer et al.~\cite{DAC99} or domain
specific pattern collections for mobile robots, are general and should
be useful for other specifications as well. Thus, to support reuse across
specifications, we have added an import mechanism to Spectra. A
specification can import other specifications to use their patterns and
predicate definitions (see Sect.~\ref{sec:predicates} and
Sect.~\ref{sec:patterns}).

\subsubsection{Syntax}
The syntax of imports is shown in Fig.~\ref{gra:import}. We extend
specifications with import statements that each start with the keyword
\key{import}, contain a filename in quotation marks, and end with a
semicolon.

\begin{figure}\hspace*{\fill}\fbox{Grammar}\vspace{-0.85pt}
\begin{lstlisting}[language=grammar,mathescape]
<|specWithImports|> replaces <|spec|>::= 
  ($\key{import}$ $\synt{"}$<|file|>$\synt{";}$)$^*$ 
  $\key{spec}$ <|name|>
  ($elems$+=<|specElem|>)$^+$
\end{lstlisting}
\caption{Grammar for imports of Spectra}
\label{gra:import}
\end{figure}

\paragraph{Well-formedness rules} The names of patterns and predicates
in all imported and in the current Spectra files must be unique. The
bodies of all (transitively) imported predicates may only contain
references to other predicates and their parameters (e.g., they may not
reference system or environment variables or defines from other
specifications\footnote{Note that these references from predicates are
allowed in general (see Sect.~\ref{sec:predicates}) and are very
convenient to use.}).

\subsubsection{Semantics}
We define the semantics of imports by a translation of imports to
Spectra without imports. Given a Spectra file with imports, we copy all
(transitively) referenced patterns of the imported files
into the current file. We also copy all (transitively) referenced
predicates into the current file and remove the import statements.

\section{Spectra Tools}
\label{sec:tools}


Spectra comes with Spectra Tools, a set of analyses and tools packaged as an
extensible set of Eclipse plug-ins.  An eclipse editor for Spectra is
implemented using XText~\cite{XTEXT}.  All symbolic representations and
operations are carried out using a BDD library (several libraries can be used,
including e.g., CUDD~\cite{CUDD}). 

Spectra Tools is available from~\cite{SpectraWebsite}, together with a user
guide and many example specifications.  We encourage the interested reader to
try it out.

We now give an overview of the different analyses and tools.

\subsection{Synthesis of Concrete and Symbolic Controllers}

First and foremost, Spectra Tools includes two synthesis features, which take the
specification as input and output a correct-by-construction controller, if one
exists.  After a series of basic checks and translations, synthesis is
performed symbolically following the algorithms described in~\cite{BJP+12},
with performance heuristics adapted from~\cite{FirmanMR17}.

The engineer can choose the form of the synthesis output.  First, the output may
be a concrete controller, which is further presented in the console or used for
application-specific code generation. As the concrete controller is complete for
the environment and deterministic for the system, i.e., in every state it
accepts all inputs from the environment and deterministically responds with an
assignment to the system variables, it allows for straightforward
application-specific code generation.

Second, the output may be a symbolic controller, following ideas from~\cite{BJP+12}.
Roughly, the symbolic controller consists of two functions, represented using
BDDs.  The first BDD describes the set of allowed initial states, while the
second describes the controller's allowed transitions.  The engineer can save it
and execute it, see below in Sect.~\ref{subsec:execution}.  

One advantage of the concrete controller output is that it enables
direct code generation.  When it is very small, it is also simple to
read and inspect.  However, it does not scale.  Its generation is slow, due to the need to
enumerate all states, and its size is beyond what any engineer can
manually inspect.
Moreover, sometimes the size of straightforward generated code prohibits
compilation. All
these disadvantages motivated us to design and implement the symbolic
controller.

\subsection{Controller Execution}
\label{subsec:execution}

Spectra Tools provides several different means to execute and simulate the synthesized
controller, concrete or symbolic.

First, code generation from concrete controllers.  The concrete controller is
directly translated to a simple implementation of a state machine. We currently
have application-specific Java code generation tailored to run on Lego NXT
robots.

Second, an execution API for symbolic controllers.  The symbolic controller is
loaded and iteratively called, at runtime, with the current inputs from the
environment, to provide the next outputs (assignment to system variables).
This runtime environment requires a BDD library, however its use is limited to
single calls to extract a satisfying assignment.  Our students have used it to
execute their Lego robots (where the actual execution runs on a Raspberry Pi)
and developed some small example standalone Java applications.

Third, a simulation environment, inside Eclipse, we call Controller Walker.  The
Walker uses the execution API for symbolic controllers.  It shows the engineer
the current state of the controller (values of environment and system variables)
and allows her to `walk', step by step, forward and backward, on its transition
system. It serves as the main tool for closely inspecting the behavior of the
synthesized controller.

\subsection{Analyses of Unrealizability}
\label{subsec:unreal}

One of the challenges of writing specifications for synthesis is unrealizability.
Spectra Tools provides several means to deal with unrealizable specifications.

First, identifying unrealizability and computing an unreazlizable core.  When
synthesis fails due to unrealizability, the engineer receives an appropriate
message.  Then, she can ask to compute an unrealizable core, i.e., a locally
minimal subset of the specification's guarantees, which already makes it
unrealizable.  Spectra Tools computes the core using DDmin~\cite{Zeller99},
following ideas described in~\cite{KonighoferHB13} and heuristics investigated
in~\cite{FirmanMR17}. The core guarantees are highlighted on the specification
(using Eclipse standard editor markers). The user guide in~\cite{SpectraWebsite}
includes example screenshots.

Second, computing a concrete counter strategy.  Given an unrealizable
specification, the engineer can generate a concrete counter-strategy, which
specifies one strategy for the environment to force any system to violate the
specification.  The counter-strategy is computed by playing a Rabin game,
following the algorithm from~\cite{KonighoferHB13,MaozS13Trace}.  The
counter-strategy can be presented in simple textual format on the console. It
can also be interactively simulated, together with the JVTS, see next.

Third, computing a JVTS.  As concrete counter-strategies may be very large and
difficult to understand, in recent work~\cite{KuventMR17} we have presented the
Justice Violations Transition System (JVTS), a symbolic representation of
a counter-strategy. The JVTS is much smaller and simpler than its corresponding
concrete counter-strategy. Moreover, it is annotated with invariants that
explain how the counter-strategy forces the system to violate the specification.
We compute the JVTS symbolically, and thus more efficiently, without the
expensive enumeration of concrete states. Finally, we provide the JVTS with an
on-demand interactive concrete and symbolic play. See~\cite{KuventMR17}.

\subsection{Analyses of Non-Well-Separation}
\label{subsec:nws}

One way a controller may satisfy a specification is by preventing the
environment from satisfying the assumptions, without satisfying the
guarantees. Although valid, this solution is usually undesired.
Specifications that allow it are called
non-well-separated~\cite{KleinP10}. In~\cite{MaozR16} we have shown
that non-well-separation is a common problem in specifications.

Spectra Tools provides means to identify and investigate non-well-separation.
The engineer can check her specification for well-separation.  If the
specification is not well-separated, information about the specific type of
non-well-separation found is displayed.  Furthermore, the engineer can ask to compute a
strategy that shows how the environment can be forced to violate its
assumptions.  Finally, a non-well-separation core, a minimal set of assumptions
that lead to non-well-separation, can be computed and highlighted on the
specification (using Eclipse standard editor markers). See~\cite{MaozR16}.

\subsection{Additional Analyses}

Spectra Tools provides several additional analyses that aim at helping engineers
write higher quality specifications.  We give some examples below.

First, some assumptions and guarantees may be trivially false or trivially true.
Clearly, such assumptions or guarantees point to a problem in the quality of the
specification, even when they are just redundant.  We provide a basic analysis
that looks for such trivial assumptions and guarantees and highlights them to the
engineer.

Second, some Spectra language elements are not easy to write correctly. 
For example, monitors contain constraints inside their body that should
only restrict the monitor variable itself. However, it is easy to write
constraints that are contradicting or constraining other variables.
These malformed monitors might lead to unrealizability of the whole
specification. Spectra Tools implements checks and highlighting for
monitors to rule out this reason for unrealizability.

Third, as one would expect from an engineer-friendly editor for writing and
reading specifications, Spectra Tools editor provides an outline, syntax
coloring, type checks, and specification completion.  These are implemented by
taking advantage of the rich XText~\cite{XTEXT} APIs.

\section{Conclusion}
\label{sec:conclusion}

The definition and development of Spectra and Spectra Tools are part of the
SYNTECH project~\footnote{SYNTECH:
\url{http://smlab.cs.tau.ac.il/syntech/}}, which aims at bridging the gap
between the theory and algorithms of reactive synthesis on the one hand and
software engineering practice on the other.  We use the language and tool set to
learn about the challenges in bridging this gap and to develop and evaluate
possible means to address them.

In this paper we presented the Spectra language.  Spectra provides means to
specify assumptions and guarantees for a reactive system, using high-level
constructs such as patterns and monitors. We further presented an overview of
Spectra Tools, a set of analyses and tools providing synthesis into
correct-by-construction controllers, means to execute and simulate the
synthesized controllers, and additional analysis aimed at helping engineers
write higher-quality specifications.  

\section*{Acknowledgements} 

We thank Elizabeth Firman, Aviv Kuvent, Or Pistiner, Rafi Shalom, Ilia Shevrin, and Yoni Wolbe
for their contribution to the implementation of Spectra Tools. 

This project has received funding from the European Research Council (ERC) under the European
Union's Horizon 2020 research and innovation programme (grant agreement No
638049, SYNTECH).

\newpage

\appendix
\section{Spectra Grammar}

We show a combination of the grammars of the Spectra kernel from
Fig.~\ref{gra:kernel} and of all extensions from
Sect.~\ref{sec:langElements} in Fig.~\ref{fig:combined}. As mentioned in
Sect.~\ref{sec:grammarNotation} this grammar includes simplifications
for readability and in order to fit it on a single page.

\begin{figure}\hspace*{\fill}\fbox{Grammar}\vspace{-0.85pt}
\begin{lstlisting}[language=grammar,mathescape]
<|spec|> ::= ($\key{import}$ $\synt{"}$<|file|>$\synt{";}$)$^*$
  $\key{spec}$ <|name|>
  (<|specElem|>)$^+$

<|specElem|> ::= <|varDec|> | 
  <|assumption|> | 
  <|guarantee|> | 
  ($\key{define}$ <|name|> $\synt{:=}$ <|exp|> $\synt{;}$) |
  ($\key{type}$ <|name|> $\synt{=}$ <|type|> $\synt{;}$) |
  ($\key{predicate}$ <|name|> $\synt{(}$ <|typedParam|> ($\synt{,}$ <|typedParam|>)$^*$ $\synt{) \{}$ <|exp|> $\synt{\}}$) |
  ($\key{monitor}$ <|type|> <|name|> $\synt{\{}$ (<|tempConstraint|>  $\synt{;}$)$^+$ $\synt{\}}$) |
  ($\key{pattern}$ <|name|> $\synt{(}$ <|name|> ($\synt{,}$ <|name|>)$^*$ $\synt{) \{}$) <|patVar|>$^+$ (<|tempConstraint|>  $\synt{;}$)$^+$ $\synt{\}}$)
  
<|typedParam|> ::= <|type|> <|name|>

<|varDec|> ::= ($\key{sys}$ | $\key{env}$) <type> <name> $\synt{;}$

<|type|>  ::= $\key{boolean}$ |
  <|name|> |
  ($\synt{\{}$ $vals$=<|name|> ($\synt{,}$ $vals$+=<|name|>)$^*$ $\synt{\}}$) |
  ($\synt{Int(}$$lower$=<|int|>$\synt{..}$$upper$=<|int|>$\synt{)}$)
  
<|assumption|> ::= $\key{asm}$ (<|name|> $\synt{:}$)$^?$ <|tempConstraint|> $\synt{;}$

<|guarantee|> ::= $\key{gar}$ (<|name|> $\synt{:}$)$^?$ <|tempConstraint|> $\synt{;}$
  
<|tempConstraint|> ::= ($\key{ini}$  | $\key{alw}$ | $\key{trans}$ | $\key{alwEv}$) <|exp|> 
   
<|exp|> ::= <|primExp|> | 
  ($\synt{(}$ <|exp|> $\synt{)}$) |
  (<|exp|> <|binaryOp|> <|exp|>) |
  (<|unaryOp|> <|exp|>)
  
  
<|unaryOp|> ::= $\synt{!}$ | $\op{next}$ | $\synt{-}$ | $\key{Y}$ | $\key{H}$ | $\key{O}$

<|binaryOp|> ::= $\synt{\&}$ | $\synt{|}$ | $\synt{->}$ | $\synt{=}$ | $\synt{<->}$ | $\synt{+}$ | $\synt{-}$ | $\synt{*}$ | $\synt{/}$ | $\synt{mod}$ | $\synt{<}$ | $\synt{>}$ | $\synt{<=}$ | $\synt{>=}$ | $\key{S}$

<|primExp|> ::= $\key{true}$ | $\key{false}$ | <|name|> | <|int|> | 
  (<|name|> $\synt{(}$ <|exp|> ($\synt{,}$ <|exp|>)$^*$ $\synt{)}$) 
\end{lstlisting}
\caption{Grammar of Spectra including the Spectra kernel from Fig.~\ref{gra:kernel} and all extensions from Sect.~\ref{sec:langElements}}
\label{gra:combined}
\end{figure}

Spectra provides more verbose alternative to most keywords shown in Tbl.~\ref{tbl:alternatives}. We have
decided to use the shorter keywords in Sect.~\ref{sec:kernel} and
Sect.~\ref{sec:langElements} as they seem to be preferred when writing
specifications.

\begin{table}
  \begin{tabular}{|l| l|}
    \hline
    \textbf{Keyword} & \textbf{Alternative}\\
    \hline
    \key{alw} & \key{always} \\
    \hline
    \key{alwEv} & \key{alwaysEventually} \\
    \hline
    \key{asm} & \key{assumption} \\
    \hline
    \key{env} & \key{input} \\
    \hline
    \key{gar} & \key{guarantee} \\
    \hline
    \key{H} & \key{HISTORICALLY}\\
    \hline
    \key{ini} & \key{initially} \\
    \hline
    \key{O} &\key{ONCE}\\
    \hline
    \key{S} & \key{SINCE} \\
    \hline
    \key{sys} & \key{output} \\
    \hline
    \key{Y} & \key{PREV} \\
    \hline
  \end{tabular}
  \caption{Overview over keywords and their alternatives (in alphabetical order)}
  \label{tbl:alternatives}
\end{table}

\clearpage

\bibliographystyle{abbrv}

\bibliography{doc} 

\end{document}